\documentclass[11pt,preprint]{aastex}
\newcommand{\eg}{{\it e.g.}}

\newcommand{\unit}[1]{\ifmmode\,{\rm #1}\else$\,{\rm #1}$\fi}

\usepackage{natbib}
\slugcomment{Accepted for Publication AJ}

\shorttitle{Central Component of Q0957+561}
\shortauthors{Haarsma et al.}

\begin{document}

\title{The Central Component of Gravitational Lens Q0957+561}

\author{
Deborah B.\ Haarsma\altaffilmark{1},
Joshua N.\ Winn\altaffilmark{2,3},
Irwin Shapiro\altaffilmark{3},
Joseph Leh{\'a}r\altaffilmark{3,4}
}

\altaffiltext{1}{Calvin College, 1734 Knollcrest SE, Grand Rapids, MI
  49546, dhaarsma@calvin.edu}

\altaffiltext{2}{Department of Physics, Massachusetts Institute of
  Technology, 77 Massachusetts Avenue, Cambridge, MA 02139}

\altaffiltext{3}{Harvard-Smithsonian Center for Astrophysics, 60
  Garden St., Cambridge, MA 02138}

\altaffiltext{4}{CombinatoRx, Inc.}

\begin{abstract}

  In 1981, a faint radio source (G$'$) was detected near the center of
  the lensing galaxy of the famous ``twin quasar'' Q0957+561. It is
  still unknown whether this central radio source is a third quasar
  image or an active nucleus of the lensing galaxy, or a combination
  of both. In an attempt to resolve this ambiguity, we observed
  Q0957+561 at radio wavelengths of 13~cm, 18~cm, and 21~cm, using the
  Very Long Baseline Array in combination with the phased Very Large
  Array and the Green Bank Telescope.  We measured the spectrum of
  G$'$ for the first time and found it to be significantly different
  from the spectra of the two bright quasar images. This finding
  suggests that the central component is primarily or entirely
  emission from the foreground lens galaxy, but the spectrum is also
  consistent with the hypothesis of a central quasar image suffering
  free-free absorption.  In addition, we confirm the
  previously-reported VLBI position of G$'$ just north of the optical
  center of the lens galaxy.  The position slightly favors the
  hypothesis that G$'$ originates in the lens, but is not conclusive.
  We discuss the prospects for further clarification of this issue.

\end{abstract}


%%%%%%%%%%%%%%%%%%%%%%%%%%%%%%%%%%%%%%%%%%%%%%%%%%%%%%%%%%%%%%%%%%%%%%%%%%
\section{Introduction}

When a galaxy lies close to the line of sight to a quasar, multiple
images of the quasar may be produced.  Almost all known lensed quasars
have an even number of images---either two or four---despite a
mathematical proof that nonsingular lens models always produce an odd
number of images \citep{dyer80a,burke81a}.  The missing image is the
``central image,'' corresponding to the central maximum in
light-travel time.  Central images are hard to detect because they are
highly demagnified by the dense core of the lensing galaxy.  They are
worth finding, however, because the properties of central images are
unique probes of the inner 10-100 parsecs of galaxies that are too
distant to resolve with ordinary observations.

Even non-detections of central images can be useful constraints on the
central structure of galaxies.  An upper limit on the flux density of a
central image corresponds to a lower bound on the central surface
density of the lens galaxy.  In the context of particular galaxy
models, the absence of a detectable central image can be used to set
the maximum size of any constant-density core
\citep{narasimha86a,wallington93a} or the steepness of any central
density cusp \citep{rusin01a,evans02a,keeton03b}.  Supermassive black
holes also affect central images: they can prevent the formation of a
central image, or produce an additional image
\citep{mao01a,bowman04a,rusin05a}. Microlensing by stars in the lens
galaxy can demagnify a central image even further
\citep{bernstein99a,dobler07a}.

Besides the faintness of the central image, there are two other
reasons why it has proven difficult to identify central images. One
reason is the close proximity between the expected position of the
central image and the center of the lensing galaxy. The expected
separation is less than $\sim$10 milliarcseconds for an isolated
lensing galaxy, although it can be larger when the galaxy is part of a
cluster that provides additional magnification
\citep{inada05a}. Resolving such a small separation would often
require space-based imaging at optical or near-infrared wavelengths,
or Very Long Baseline Interferometry (VLBI) at radio wavelengths.  A
second and related reason is that even when a central point source is
detected, it is difficult to decide conclusively whether it is a
central image of a background quasar, or an active galactic nucleus
(AGN) in the lens, or a combination of these two possibilities. This
decision is difficult not only because the point source is faint and
challenging to characterize, but also because propagation effects may
alter the properties of the central image and cause it to appear
different from the other lensed images, thereby interfering with the
usual tests for lensed image identification. Such propagation effects
include extinction by dust at optical wavelengths, and scintillation and
free-free absorption at radio wavelengths; these effects are expected
to be more important for lines of sight passing very near the center
of the foreground galaxy, where on average the density of dust and
plasma should be greatest.

A central image was detected for the bright optical lens
APM~08279+5255 \citep{ibata99a,egami00a,munoz01a}, but it seems
probable that this image is the result of a highly elongated mass
distribution rather than the high central density of the lensing
galaxy \citep{lewis02a}.  More recently, \cite{winn04a} presented
evidence for a central image in PMN~J1632--0033, thereby setting an
upper limit on the mass of any central black hole and a lower limit on
the central surface density of the lens galaxy. A central image has
also been detected at optical wavelengths for SDSS~J1004+4112
\citep{inada05a}.  In addition, there are cases in which central radio
components have been detected but are most likely to be the active
nucleus of the lens galaxy \citep{chen93a,fassnacht99b}, and cases
where recent work has placed stringent upper limits on the flux
density of any central component \citep{boyce06a,zhang07a}.

This paper is concerned with the very first lensed quasar to have been
discovered, Q0957+561 \citep{walsh79a}. More than 20 years ago, a
central radio component was detected in this system
\citep{gorenstein83a}, but there is still a lingering uncertainty
about its interpretation.  The system consists of a quasar at redshift
1.41 that is gravitationally lensed into 2 images (A and B) by an
elliptical galaxy G1 and a surrounding set of galaxies mostly at
redshift 0.36 \citep{stockton80a,young81a}. Quasar images A and B
appear as point sources in optical images and as core-jet sources in
radio images. Observations with the Very Large Array (VLA) reveal a
radio source, dubbed G, close to the center of the
optically-identified lens galaxy G1 \citep{greenfield80a,roberts85a}.
Radio observations with VLBI have revealed an unresolved point source
near the center of G1 \citep{gorenstein83a}. The VLBI source is known
by the different name G$'$ because its relationship to G is
unclear.  The VLA and VLBI observations are sensitive to very
different spatial scales of radio emission; typically, VLBI
observations have an angular resolution 10-100 times finer than VLA
observations and are blind to most structures that are resolved by the
VLA.

Modelers of this lens have typically assumed that G$'$ marks
the center of the lensing galaxy, and that any central quasar image is
undetectably faint \citep[see,
e.g.,][]{barkana99a,bernstein99a,chae99a}. If instead G$'$ is the
central image, then the lens galaxy has a shallower central
gravitational potential than has been assumed by those modelers.
Traditionally this identification has not been a major concern,
because the lens modeling literature has been preoccupied with
the connection between the Hubble constant and measurements of the
A--B time delay, which is not greatly affected by the innermost
$\sim$100~pc of the galaxy mass distribution. However, as noted above,
the interpretation of G$'$ does have an impact on our understanding of
the mass distribution of the central regions of elliptical galaxies.

%Barkana99: assume optical G1 for the lens galaxy position.  Follow
%gorenstein94a: if G$'$ is the lens, then a 5sigma limit that central<B/30.
%
%Bernstein99a: G1 so close to G$'$, so assume G$'$ is center of lens potential.
%Don't apply constraint of lack of detection of 3rd image, since 
%shape of inner potential has little effect on Ho
%
%chae99a: use G$'$ for optical position, but G1 vs. G$'$ positions make
%little difference.  C/B limit is quite small [assuming G$'$ is lens], so
%any model with a small core radius satisfies it

We undertook new radio observations of Q0957+561 in an attempt to
clarify the nature of G$'$.  Our strategy was to compare the radio
spectral-energy distributions of A, B, and G$'$ over as large a range
of radio wavelengths as feasible.  For all practical purposes, gravitational
lensing is wavelength-independent. Thus, the central image would have
the same spectrum as images A and B, were it not for the confounding
factors of intrinsic variability (in conjunction with different time
delays for the different images), the magnification gradient along the
jet images (in conjunction with different spectral energy
distributions for the core and jets), and propagation effects specific
to the line of sight of each image (such as scintillation and
free-free absorption), which must be taken into account. An AGN in the
lens would not have the same spectrum as the quasar images except by
coincidence.  Although the Q0957+561 system has been studied
extensively at radio wavelengths, flux-density measurements of G$'$ have not
been well documented.  Previously, the only flux-density measurement
with a reported uncertainty was by \cite{gorenstein83a} at 13~cm. The
VLA source G has been documented more completely, including
measurements at wavelengths of 2~cm, 3.6~cm, and 6~cm by
\cite{harvanek97a}.

% We actually have Campbell95, Campbell94 6cm data from 1987-1993,
%  but would have to be reanalyzed to extract G' (was it even correlated
%  near G'?

In the following section, we describe our VLBI observations of this
system and G$'$ in particular. We describe the data reduction
procedures in \S\ref{reduction}, and we present the empirical results
in \S\ref{results}. In \S\ref{discussion} we discuss the spectrum, and
in \S\ref{future} we review the prospects for further progress on this
issue.

%%%%%%%%%%%%%%%%%%%%%%%%%%%%%%%%%%%%%%%%%%%%%%%%%%%%%%%%%%%%%%%%%%%
\section{Observations}
\label{obs}

We observed Q0957+561 with VLBI at wavelengths of 13, 18, and 21~cm.
We made the 13~cm (2.27~GHz) observations over 14 hours on UT 1997
February 13, using the 10 antennas of the VLBA; the 18~cm (1.67~GHz)
observations over 14 hours on UT 1997 March 19-20, using the 10 VLBA
antennas and also the phased VLA and the Green Bank 140-foot telescope
(since decommissioned); and the 21~cm (1.43~GHz) observations over 4
hours on UT 2005 December 30, using the 10 VLBA antennas and also the
phased VLA and the Green Bank 100~m telescope. We attempted a fourth
observation at 1.3~cm (22~GHz) on UT 2005 November 23, but we did not
detect fringes on either the target or the chosen calibration source
(J0957+5522); we do not discuss those data further.

To calibrate the relative antenna gains, we switched between Q0957+561
and a nearby, bright, and fairly compact source, J1035+5628.  We used
a cycling time of 5.5 minutes for the 13~cm observations, including 4
minutes spent on Q0957+561. At 18~cm we used a cycling time of 7
minutes, including 5~minutes spent on Q0957+561. Based on the results
of these observations, we found that self-calibration was sufficient
to calibrate the relative antenna gains, and hence for the 21~cm
observations we switched to a calibration source less frequently; we
visited the calibration source J0929+5013 every 15 minutes.
  
For all of the observations, we divided the observing bandwidth of 32
MHz per polarization into 4 sub-bands. Both senses of polarization
were recorded with 2-bit sampling. The data were correlated in
Socorro, New Mexico. For the 13 and 18~cm observations, we used
successively three different correlation centers, chosen to match the
expected locations of components A, B, and G$'$. For the 21~cm
observations, we used two different correlation centers, chosen to
match the expected locations of component A and of the midpoint between
B and G$'$.  In each case the correlation produced 16 channels of
width 500 kHz from each sub-band, with an integration time of 4
seconds.

%%%%%%%%%%%%%%%%%%%%%%%%%%%%%%%%%%%%%%%%%%%%%%%%%%%%%%%%%%%%%%%%%%%
\section{Data Reduction}
\label{reduction}

We used standard procedures within the NRAO Astronomical Image
Processing System (AIPS) for calibration, as summarized here. We
discarded obviously corrupted data and data taken at elevations less
than 15 degrees.  We calibrated the visibility amplitudes using
on-line measurements of antenna gains, system temperatures, and
voltage offsets in the samplers.  We removed large delay errors by
fringe-fitting the data from a 1.5-minute observation of a bright
calibration source at each wavelength and applying the delay
corrections to all the data at that wavelength. We corrected residual
rates, delays, and phases by fringe-fitting the data from J1035+5628
or J0929+5013, with a 2-minute averaging time. After calibration, we
averaged the data in time and in frequency to reduce the data volume
while still preserving a fine enough degree of sampling to limit the
smearing to a few per cent over the desired field of view of a few arc
seconds.

We also performed self-calibration with AIPS. For each wavelength
band, we created a preliminary map of the region around image A,
using the CLEAN algorithm.  We then self-calibrated the gain phases
with reference to the CLEAN model.  We repeated the process of CLEAN
and self-calibration with progressively shorter averaging times until
no further improvement was noted.  We also attempted self-calibration
of the antenna amplitudes, but only for the 18~cm data did this result
in a significant improvement.

While creating our final maps, we experimented with various schemes
for weighting the visibility data.  In the absence of calibration and
deconvolution errors, ``natural'' weighting based only on the expected
thermal noise of each antenna should lead to a lower noise level than
any other weighting scheme. This was indeed the case for the 13~cm
data, which were taken with a homogeneous array of the 10 VLBA
stations. However, for the 18~cm and 21~cm data, we found the best
results were obtained with the AIPS parameter ROBUST~$=$~0
\citep{briggs95a}, which is a compromise between natural weighting and
``uniform'' weighting (in which corrections are applied to give equal
weight to each point in Fourier space).  Apparently, the dynamic range
of the 18~cm and 21~cm observations was limited to a few hundred by
calibration and deconvolution errors, preventing the improvement in
sensitivity from the inclusion of the larger antennas from being fully
realized.  In particular, for the 21~cm data, the thermal noise limit
is calculated to be 20~$\mu$Jy~beam$^{-1}$ for a naturally weighted
map, and 32~$\mu$Jy~beam$^{-1}$ for a map with ROBUST~$=$~0. In
contrast, our best map (obtained with ROBUST~$=$~0) has an rms noise
level of 45~$\mu$Jy~beam$^{-1}$. The best indication that the problem
is indeed dynamic range comes from the Stokes $Q$ map, in which the
peak flux density is much smaller than in the $I$ map and the noise
level is much closer to the theoretical limit. Similar conclusions
based on similar data have been reached by \cite{boyce06a} and
\cite{zhang07a}. For the final maps, we simultaneously deconvolved 
the fields around both A and B, to prevent sidelobes from image B
from lowering the dynamic range in the vicinity of A, and vice versa.

To characterize the jet structures of quasar images A and B, we also
created a set of maps at 18~cm using uniform weighting of the
visibility data and a circular restoring beam of radius 4~mas after
deconvolution. These maps are shown in Figure~\ref{fig.18cm4mas} (and
were previously reported by \citealp{haarsma00a}). They have rms noise
of 55 (67)~$\mu$Jy~beam$^{-1}$ in the vicinity of A (B). The
ROBUST~$=$~0 maps had a lower noise level in presumably blank regions
of the sky, but for jet characterization we prefer the uniformly
weighted maps because uniform weighting provides a better suppression
of the sidelobes. The circular restoring beam also allows for a visual
comparison to previous work.

%%%%%%%%%%%%%%%%%%%%%%%%%%%%%%%%%%%%%%%%%%%%%%%%%%%%%%%%%%%%%%%%%%%%%%%
\section{Results}
\label{results}

% shape/jet
We successfully detected G$'$ at 13~cm, 18~cm, and 21~cm. We measured
the flux density of G$'$ by fitting a point-source model to the final
maps (using the AIPS task JMFIT). The measured component flux
densities and map rms noise are given in Table~\ref{table.fluxes}.
Fitting G$'$ with a two-dimensional Gaussian function did not change
the results significantly, indicating that G$'$ was unresolved by our
observations. Thus there is no evidence for a jet or any additional
structure in G$'$ larger than the size of the synthesized beam. The
principal axes of the synthesized elliptical beam (in milliarcseconds)
and the position angle of the major axis (east from north) at 13~cm,
18~cm, and 21~cm are $5.8\times 4.3$~$(-13\arcdeg)$, $7.4\times
5.5$~$(-11\arcdeg)$, and $14\times 5.2$~$(1\fdg7)$, respectively.  In
addition, there are no sources in the $1.2\arcsec \times 1.2\arcsec$
region around G$'$ brighter than 4$\sigma$ above the noise.
% (see Figure~\ref{fig.G}).

% AB fluxes
We also measured the total flux densities of A and B at each epoch,
using the sum of the flux density within an aperture that encompassed
the bright point source and the jet (using the AIPS task
TVSTAT). These results are also given in
Table~\ref{table.fluxes}. There is some subjectivity involved in
defining the aperture that encompasses the entire jet; based on the
results from apertures of different sizes and shapes, we find the
uncertainty in the total flux densities of A and B to be about 5\%.

% variability
At 13~cm, we measured the flux density of G$'$ to be
$0.83\pm0.07$~mJy.  \citet{gorenstein83a} measured the 13~cm flux
density of G$'$ to be $0.60\pm 0.10$~mJy based on data obtained in
1981, an observation sixteen years earlier than ours.  The difference
in the two results is $0.23\pm 0.12$~mJy, suggesting that G$'$ is
variable, but this difference may also be due to the two sets of VLBI
data being from different arrays and being calibrated by different
means. 

% position
The coordinates we measured for G$'$ at each wavelength agree with
each other and with the coordinates reported by \cite{gorenstein83a}
within 1$\sigma$; hence, we confirm the previously reported
position. A compilation of reported coordinates for G$'$, G, and G1 is
given in Table~\ref{table.Gpos} and plotted in
Figure~\ref{table.Gpos}. Notably, the most precise positions of G$'$,
G, and G1 are in disagreement at the 2-3$\sigma$ level, and in
particular G$'$ is located north of the optically detected lens galaxy
G1.

% AB jet
The jets of A and B shown in Fig.~\ref{fig.18cm4mas} at 18~cm appear
similar to those presented by \cite{garrett94a} (hereafter G94), but
with some intriguing differences described below.  Due to the
difficulties described in \S\ref{reduction} in achieving the thermal
noise limit, these maps do not have a significantly better
signal-to-noise ratio than the G94 maps. Moreover, the component most
significant for modeling (component 5 of G94, the brightest part of
the jet), did not change significantly in location, brightness, or
position angle in the 8 years between the G94 observations and our
observations. Thus, the new maps are unlikely to provide significant
improvement in the magnification matrix of the lens, and hence we have
not attempted to provide a comprehensive quantitative description of
the jet maps.

\cite{campbell95a} monitored the jets from 1987 to 1993 with an
angular resolution of about 1~mas and did not detect any motion of the
components along the jets. Our jet maps do, however, suggest some
small but interesting changes in the jets between 1989 and 1997
(marked on Figure~\ref{fig.18cm4mas}). These differences are not much
more significant than spurious noise features, but they could suggest
superluminal motion of the jet components.  In A and somewhat in B,
the region 5-20~mas from the core is somewhat brighter in 1997 than in
1989, perhaps as component 2 of G94 moved further from the core.  In
A, the region about 65~mas from the core (component A6 in G94) is
brighter, and in B, a new component has appeared on the end of the
jet.  These slight changes are not definitive enough to be useful, but
we do caution lens modelers to account for possible superluminal
motion when calculating uncertainties in the magnification matrix, as
noted by \cite{barkana99a}.

%%%%%%%%%%%%%%%%%%%%%%%%%%%%%%%%%%%%%%%%%%%%%%%%%%%%%%%%%%%%%%%%%%%%%%%
\section{The Spectrum of G$'$}
\label{discussion}

Our goal was to clarify the origin of G$'$ by comparing the broad-band
radio spectrum of G$'$ with that of A and B, over as wide a range of
wavelengths as possible. As noted in the introduction, gravitational
lensing itself is achromatic, leading to the expectation that lensed
images have the same dependence of radio flux density on wavelength,
with the caveats that intrinsic variability, magnification gradients
across the source, and propagation effects are confounding factors.

We examine the flux-density ratios G$'$/B and A/B, rather than the
flux densities themselves, because the ratios are not subject to the
extra uncertainty introduced by the overall flux-density calibration
when comparing results from different epochs. The solid symbols
plotted in Fig.~\ref{fig.Gspec} represent the VLBI measurements from
this work as well as the 13~cm VLBI measurements by
\cite{gorenstein83a}.  (The open symbols are discussed below.)  The
top panel shows the ratio A/B, while the bottom panel shows G$'$/B; in
both cases the axes are logarithmic, with identical horizontal axes
and vertical axes that both span a factor of 10.  Clearly, G$'$/B
decreases significantly with wavelength.  If the radio spectrum of
each component is described by a power law, $S_\nu \propto
\nu^\alpha$, then a least-squares fit of a straight line on the
logarithmic plot shows that the exponents for G$'$ and $B$ differ by
$\Delta\alpha = -1.04\pm 0.23$.  At face value this suggests G$'$ has
a different spectrum from B and is therefore not the central quasar
image. A proper interpretation, however, requires consideration of
several additional factors.

One factor is the intrinsic variability, coupled with the time
delays. The quasar is known to be variable in luminosity at 6~cm and
3.6~cm by as much as 30\% \citep{lehar92a, haarsma97a, haarsma99a}.
This variation is slow compared to the A-B time delay, varying by at
most 15\% over 1.1 years.  Because of the time delays, each image
displays the quasar in a different luminosity state, and thus the
instantaneous flux-density ratio may vary.  For images A and
B at 3.6~cm, this variation of A/B could be as large as $\pm15$\%.
This effect could account for some of the 30\% difference between our
1997 measurement of the 13~cm A/B flux density ratio and the 1981
measurement of the same quantity by \cite{gorenstein83a}. The
measurements of A/B at other wavelengths, observed on a variety of
dates, are consistent within $\sim$20\% across the spectrum.

The time delay between the central image and the B image is expected
to be much shorter than the delay between the A and B images, since
the time delay depends on the image positions roughly as
$\Delta\tau_{BA} \propto r_B^2 - r_A^2$ \citep{kochanek02a}, where $r$
denotes the angular distance from the image to the lens center. If
G$'$ is the central image, its distance from the lens is less than
$0\farcs01$, while A and B are at $r_A\sim5''$ and $r_B\sim1''$,
respectively.  Since the time delay between A and B is $\sim1.1$ years
\citep{haarsma97a, kundic97a, oscoz97a, schild97a}, the delay between
B and G$'$ would be less than one month. At radio wavelengths, the
most rapid observed variation of the quasar has been $\sim$2\% per
month, and the typical variation is even slower
\citep{haarsma99a}. Since the measurement uncertainty for the flux
ratio is $\sim$4\%, variability should not significantly affect our
interpretation of the spectrum of G$'$/B.

Another factor is the gradient in magnification along the extent of
the VLBI jets (\eg\ \citealp{garrett94a,barkana99a}), coupled with the
different radio spectra of the core and jets.  These factors will
cause the sum of the core and jet flux densities to vary with
wavelength in a manner specific to each quasar image
\citep{conner92a}.  Thus, the flux ratio of the two images will appear
to be wavelength dependent if the core and jet are combined.  For
Figure~\ref{fig.Gspec}, we calculated the flux ratios using the sum of
the core and jet flux density for each image, because the jet of the
central image would be less than 1~mas long and unresolved by our
measurements.

%If the radio spectrum of each component is described by a power law,
%$S_\nu \propto \nu^\alpha$, than the spectral index $\alpha$ can be
%found from observations made at two different frequencies on nearly
%the same date. Based on observations near October 1989 at 18~cm
%\citep{barkan99a} and 6~cm \citep{campbell94a}, the spectral index of
%the jet is $\alpha_J=-0.61$ and the core is $\alpha_C=-0.35$.

How much would the magnification gradient affect the G$'$/B spectrum?
The magnification of the central image and its jet depends on the
unknown central mass distribution of the lens.  Many other key
parameters are known, however, or can be derived from observations,
such as the spectral indices and magnifications of the core and jet in
the A and B images.  Assuming that the flux density of G$'$ is due
entirely to the central image and its jet, we tested a range of values
for the ratio of the G$'$ jet magnification divided by the G$'$ core
magnification.  This ratio is expected to be greater than 0.8 in
simple models; we tested a wide range of values from 0.8 to 1.6.
Holding the 18~cm flux densities fixed, we found the difference between the
G$'$/B ratio at 21~cm and the G$'$/B ratio at 2~cm to be altered by less
than 10\%.  Thus, it is unlikely that the magnification gradient is
introducing variations in the G$'$/B spectrum at a level above 10\%.

A final factor that may affect the G$'$/B spectrum is the possibility
of differential propagation effects.  G$'$ may suffer more from
propagation effects due to its proximity to the center of the lens.
Whether G$'$ is radiation from the lens or the quasar, its measured
flux density may be affected by free-free absorption and/or
scintillation due to ionized material near the lens's center.
Therefore we must consider the possibility that G$'$ is at least in
part a central image whose spectrum has been altered by propagation
effects. This issue can be addressed by checking for the
characteristic wavelength dependence of these effects (for example,
the optical depth due to free-free absorption is expected to vary
roughly as $\lambda^2$), or preferably, by observing at short enough
wavelengths that propagation effects are expected to be negligible, as
done by \citet{winn04a} for PMN~J1632--0033.  Since our 1.3~cm VLBI
observations that were intended for this purpose did not succeed,
below we compare our results with previous VLA observations at shorter
wavelengths.

As mentioned above, VLA observations revealed a central radio source G
that is located near the center of the optically detected galaxy
G1. Component G is unresolved by the VLA, save for faint north-south
extensions \citep{roberts85a,harvanek97a} reminiscent of a core and
two-sided jet of a low-luminosity AGN.  Comparisons of the flux
density of G and G$'$ must be made with caution because of the very
different spatial scales involved; the VLA observations had a beam
size of 300-1000~mas as compared to the 3-10~mas beam size of the VLBI
observations. To the VLA, G$'$ is unresolved and its flux density must
contribute fully to the total measured flux density of G. However, the
flux density of G$'$ as measured by the VLBI arrays does not include
any of the structure that has been resolved by the VLA, and is also
missing an unknown fraction of the flux density of G that is
unresolved by the VLA but that is resolved out on even the shortest
baselines of the VLBI array.

\cite{harvanek97a} decomposed G into three subcomponents: an
unresolved point source, a northern extension, and a southern
extension. The two extensions were referred to as ``remnants.''
\cite{harvanek97a} provided estimates of the flux density of each of
the three subcomponents at 6~cm and 3.6~cm.  At 2~cm, they did not
detect the remnants and provided only the flux density of the
unresolved subcomponent.  In the absence of shorter-wavelength VLBI
data, we assumed for present purposes that the unresolved portion of G
is the appropriate comparison to G$'$.  The open symbols in
Fig.~\ref{fig.Gspec} show the measurements by Harvanek et al.~(1997).
In the lower panel, we set the error bars equal to the quoted flux
density of the brighter of the two remnants, since there is likely to
be a strong correlation in the measured flux densities of each of the
three subcomponents.

Unfortunately, the short-wavelength VLA measurements of G$'$/B do not
resolve the issue.  The data are consistent with a model of free-free
absorption (dotted line in Fig.~\ref{fig.Gspec}) in which G$'$/B
varies as $\exp(-\tau_\nu)$ with $\tau_\nu \propto \nu^{-2.1}$
\citep{mezger67a}.  A least-squares fit has a $\chi^2$ of 7.7 and 6
degrees of freedom.  The data are also consistent with a model in
which G$'$ is the lens AGN with a different power-law spectrum than B
and no free-free absorption (solid line in Fig.~\ref{fig.Gspec}).  The
best least-squares fit of a straight line to G$'$/B has a $\chi^2$ of
6.4 and 6 degrees of freedom.  Thus, according to this analysis, we
cannot rely on the current spectrum to distinguish between the two
models, especially in light of the possible systematic errors 
in the G/G$'$ correspondence.

%%%%%%%%%%%%%%%%%%%%%%%%%%%%%%%%%%%%%%%%%%%%%%%%%%%%%%%%%%%%%%%%%%%%%%%
\section{Future Prospects for Understanding G$'$}
\label{future}

Although we have measured the spectrum of G$'$ for the first time, the
source of G$'$ remains ambiguous.  In this section, we review the
relevant observations relating to G$'$, and we consider the prospects
for clarifying the nature of G$'$ with future observations.  Clearly,
one way forward would be to fill in the short-wavelength part of the
spectrum with VLBI observations with a high signal-to-noise ratio at
several wavelengths.  In addition, several other observable properties
of G$'$ could be useful, namely, its overall brightness, position,
radio structure, and variability. We treat each of these in turn.

{\it Overall brightness.}---It has long been argued that G$'$ is
unlikely to be a central quasar image based on its brightness.  Its
measured flux density is reasonable for a low-luminosity AGN
\citep{roberts85a} but brighter than simple lens models would predict
for the central image.  For example, assuming that the central mass
distribution of G1 is approximately isothermal, the central image C
should be demagnified by approximately $r_C/r_E$ relative to the
bright quasar images, where $r_C$ is the distance from the central
image to the lens center and $r_E$ is the Einstein radius.  Since
$r_C$ is likely to be less than 10~mas and $r_E$ is roughly 3000~mas,
one would expect demagnification by a factor of 300 rather than the
factor of 30-50 we measure for G$'$.  (Note that if G$'$ is a
combination of the central image and the lens galaxy nucleus, they
must be separated by a fraction of a milliarcsecond to appear
unresolved in our data, and at such a separation the central image
would be even more strongly demagnified.)  However, apart from the
uncertainty in the position of G1 relative to G$'$, we are reluctant
to attach much weight to this argument because it involves some
circular logic: after all, it is the mass distribution of the lens
galaxy that we are trying to study. The order-of-magnitude argument
given here also does not take into account the mass-sheet degeneracy
\citep{falco85a} and other inherent limitations of lens modeling.

\begin{figure}
\plotone{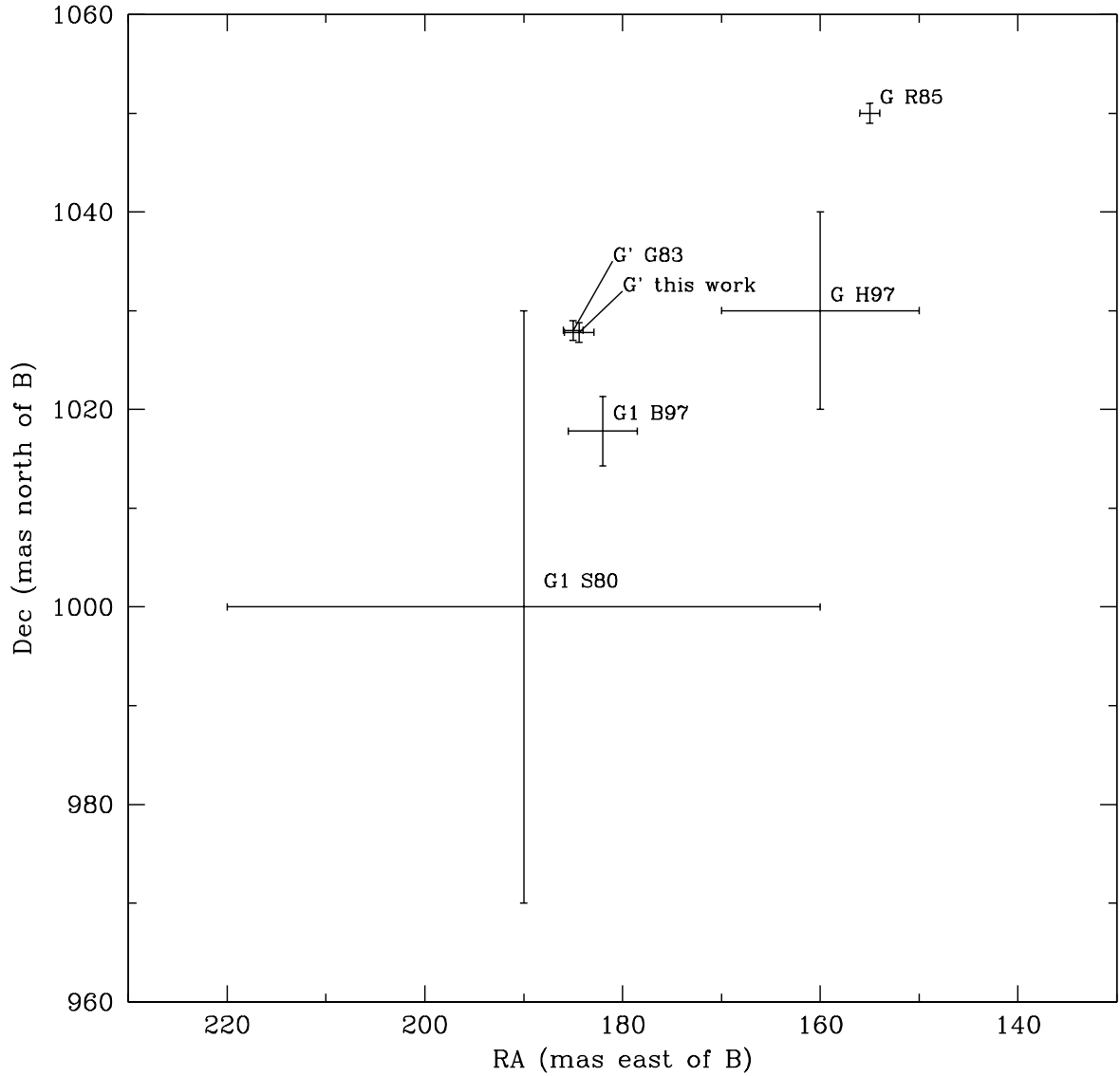}
\caption{Positions of optical lens galaxy G1, radio VLA source G, and
radio VLBI source G$'$ (see Table~\ref{table.Gpos} for references).
Note that \cite{roberts85a} indicate that the uncertainty in their
position is underestimated; the \cite{harvanek97a} uncertainties are
more typical for positions found with the VLA.
\label{fig.Gpos} }
\end{figure}

\begin{figure}
\plotone{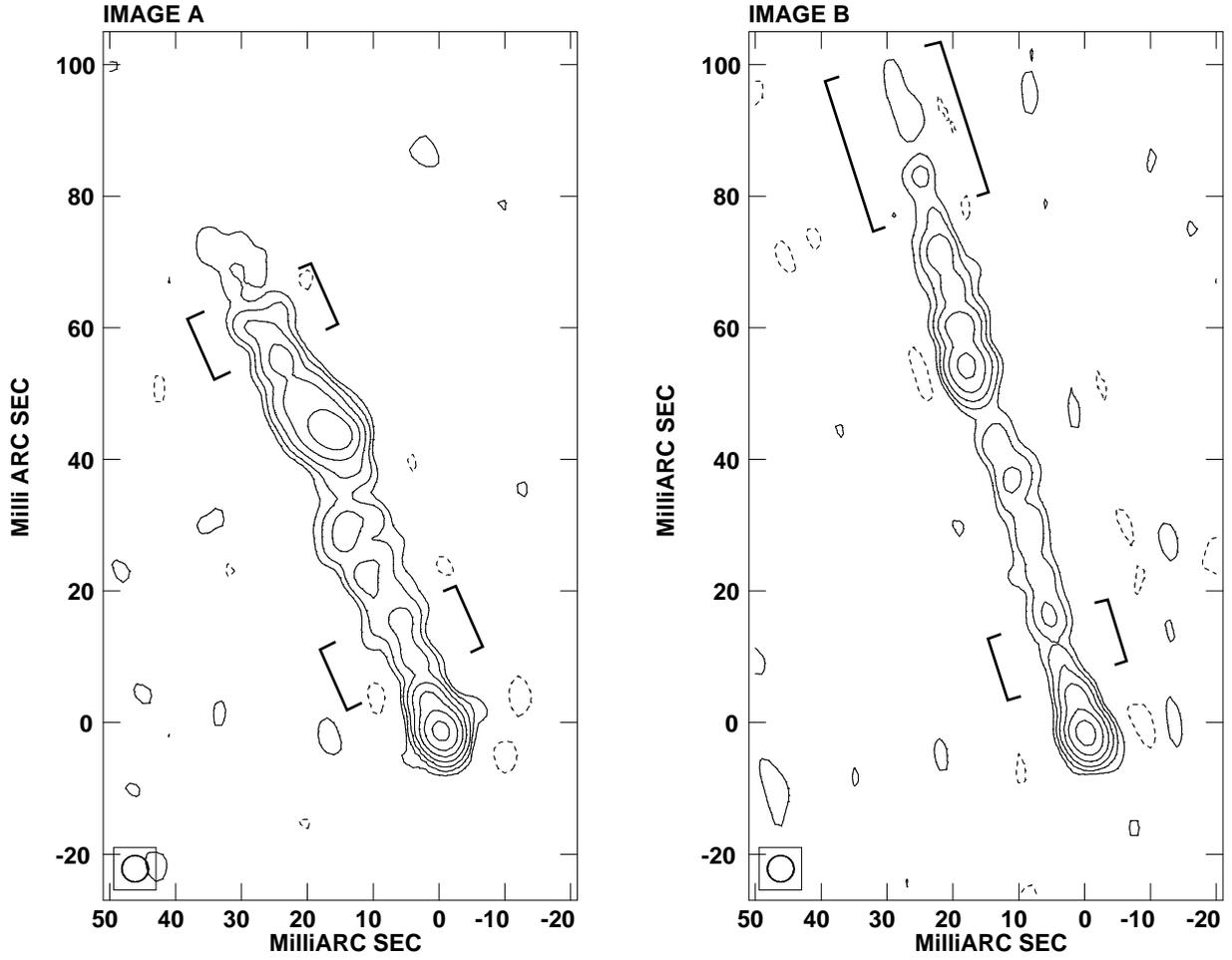}
\caption{Maps of images A and B at 18 cm as observed on
1997~Mar~19-20.  The coordinates of map A (B) are centered at J2000
10:01:20.69 +55:53:55.6 (10:01:20.84 +55:53:49.6).  The contours are
at flux densities of -0.15,0.15, 0.3, 0.6, 1.2, 2.4, 2.8, 9.6
mJy~beam$^{-1}$.  The 4~mas circular restoring beam is shown in the
lower left.  Brackets indicate regions that appear brighter in 1997
than in the 1989 observations by \cite{garrett94a}.
\label{fig.18cm4mas} }
\end{figure}

\begin{figure}
\plotone{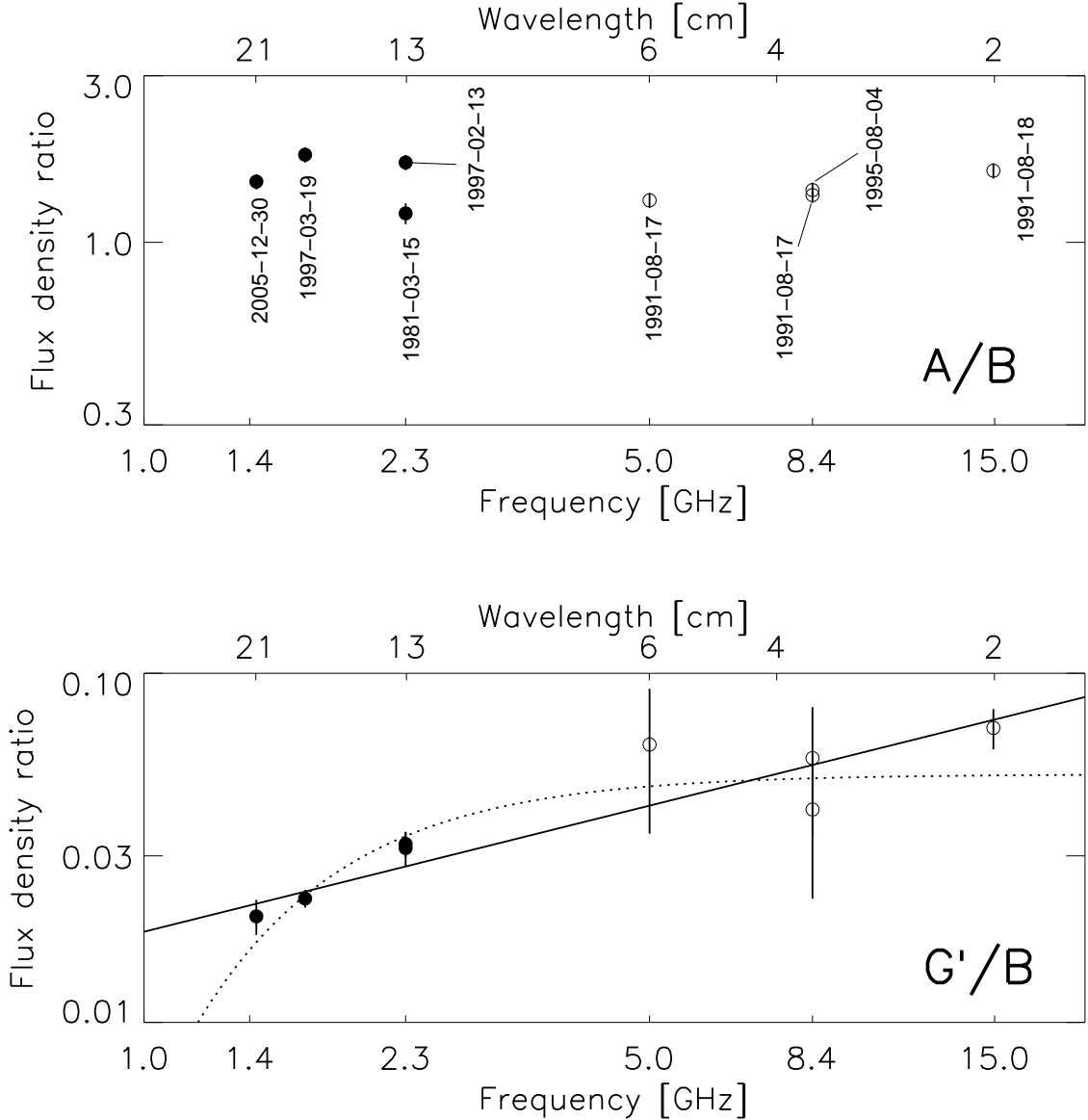}
\caption{ The flux-density ratios A/B (top panel) and G$'$/B (bottom
panel) as a function of observing wavelength. The observing dates are
displayed in the top panel. The data points at wavelengths shorter
than 10~cm (open symbols) are for the VLA component G measured by
\cite{harvanek97a}, while those points at longer wavelengths (filled
symbols) are for the VLBI component G$'$ measured in this work and at
13~cm by \cite{gorenstein83a}.  In the top panel, A/B is seen to vary
significantly, which can be at least partially attributed to intrinsic
variability and the differential time delay (see \S~5).  In the bottom
panel, the dotted curve is the best fitting 2-parameter model in which
G$'$ is a central image suffering free-free absorption ($\tau_\nu
\propto \nu^{-2.1}$) in the lens galaxy.  The solid line is the
best-fitting 2-parameter model in which G$'$ is an AGN in the lens
galaxy with a different power-law spectrum than that of the quasar
images.
\label{fig.Gspec} }
\end{figure}

{\it Position.}---Most realistic lens models would predict that the
central image should be located closer to image B than to image A.  In
such cases, it should fall south of G1, roughly along the line joining
G1 and B.  In contrast, an AGN of the lens galaxy is expected to lie
very close to the center of G1, although there is no guarantee that a
central black hole marks the precise center of mass.  Still, there is
the potential to distinguish between these possibilities with a very
accurate measurement of the G$'$--G1 angular separation and position
angle.  In this work, we confirm the \cite{gorenstein83a} VLBI
position of G$'$.  The most precise optical position of G1 comes from
{\it HST}\, observations \citep{bernstein97a} (see
Table~\ref{table.Gpos} and Figure~\ref{fig.Gpos}).  The separation
between these measurements of G$'$ and G1 is 10.6~mas, about three
times larger than the measurement uncertainty (3.6~mas, dominated by
uncertainty in the optical centroid of G1).  A 3$\sigma$ measurement
is not precise enough to conclusively confirm or deny that G$'$ and G1
are coincident.  The position angle, however, is more suggestive.
G$'$ lies to the {\it north} of G1, contrary to the expectation that
the central image would appear south of the lens.  Thus, the position
information slightly favors the hypothesis that G$'$ is the AGN of the
lens galaxy.

It is not clear whether future {\it HST}\, data or other data will
allow for a significant improvement, but measuring the position of G1
more accurately is a potentially promising avenue for future progress.
The detection of an optical point source displaced from
the center of G1 would be an excellent candidate for the optical
counterpart to G$'$ and good evidence for a central image, just as it
is for the lens system SDSS~J1004+4112 \citep{inada05a}.  The existing
{\it HST}\, data show a power-law surface brightness distribution with
no evidence for an additional central point source down to
$\sim0\farcs1$.  The non-detection of an optical counterpart to G$'$,
even at a level smaller than 1\% of A, does not rule out the
central-image hypothesis because the optical central image can be made
much fainter than the radio central image due to microlensing by stars
or extinction by dust.

{\it Radio structure.}---If G$'$ is a third quasar image, then it is
demagnified by a factor of 30-50 relative to images A and B. Since A
and B are observed to have jets that extend out to $\sim$50~mas in
VLBI maps, the corresponding jet in G$'$ is expected to have an extent
of $\sim$1~mas. This is a factor of a few smaller than the synthesized
beam of our VLBI observations, which means that the pointlike
appearance of G$'$ in our maps is not constraining.  \cite{rogers88a}
reported a core-jet structure for G$'$, but this finding is not
confirmed by our study, nor by any published study we are aware
of. Furthermore, although the detection of a slight extension in G$'$
would certainly be interesting, one would want to see the detailed
correspondence between that jet and the jets of A and B before
declaring G$'$ to be the central image. Otherwise, it could represent
a jet from the AGN in the lens, or the combination of a central image
and an AGN. This test would require considerably greater angular
resolution and sensitivity.

The VLA component G, in contrast, is certainly resolved on
$\sim$500~mas scales \citep{harvanek97a}, which is much larger than
the lensing expectation. Furthermore, investigators have observed
extended, low-surface-brightness radio lobes about 15\arcsec\ to the
north and south of G1, oriented along nearly the same direction as the
extension of G \citep{avruch97a, harvanek97a}.  These seem like clear
indications that G must include at least some contribution from an AGN
in the lens galaxy G1.  \cite{harvanek97a} noted some problems with
this interpretation, mainly that the morphology of G1 would not be
that of a typical radio galaxy of either of the two Faranoff-Riley
categories. They raised the possibility that at least some of the
extended structure near G actually originates from image B and
represents the counter-jet to the long and prominent arcsecond-scale
jet of image A, but this interpretation is also problematic. In any
case this issue is somewhat separate from the issue of the identity of
the VLBI component G$'$, except insofar as the knowledge that G1 is
definitely an active galaxy would raise the likelihood that G$'$ is
its radio nucleus.

{\it Variability.}---The background quasar is known to be variable;
thus the central image C should also vary and have the correct time
delay relative to the A and B images (although the B-C time delay
would be much shorter than the A-B delay, as discussed above).  Source
G was not seen to vary significantly in our VLA monitoring
\citep{lehar92a,haarsma97a,haarsma99a} and G$'$ has had no consistent
observations to detect variation.  Even if G$'$ were shown to vary,
that would not conclusively prove it is the central image, since the
radio core of the lens could also be variable.  The observation of
correlated variability would be required to show that G$'$ is the
central image. Since A and B are at least occasionally variable at the
30\% level, one would want to detect variations of G$'$ at that level
with a high signal-to-noise ratio.

Such an undertaking is probably too demanding on current resources to
be feasible, but it will be feasible with the next generation of radio
observatories.  As an example we consider the {\it Expanded Very Large
  Array} (EVLA), an effort to improve almost all of the relevant
observing parameters of the VLA by a factor of 5-20.  With the VLA, G
can be detected at 2~cm with a signal-to-noise ratio of 8 in a 6~hr
observation \citep{harvanek97a}.  With a factor of $\sim$10
improvement in the 2~cm band, the EVLA could obtain a signal-to-noise
ratio of 20 in about 4~hr (if the dynamic range is not as limited as
we experienced with a heterogeneous VLBI array). This measurement
could be repeated every few weeks for a few years to search for
correlated variability, although the configuration changes of the VLA
would complicate the actual observing plan. Of course the whole
strategy depends on the assumption that at 2~cm the flux density of G
is dominated by that of G$'$, an assumption that can be checked in
future VLBI observations.  Further developments in high-sensitivity
VLBI, along with the advent of the Atacama Large Millimeter Array and
eventually the Square Kilometer Array can also be brought to bear on
this stubborn problem.

%%%%%%%%%%%%%%%%%%%%%%%%%%%%%%%%%%%%%%%%%%%%%%%%%%%%%%%%%%%%%%%%%%%%%%%%%%
\acknowledgments

D.H.\ acknowledges the support of Calvin College and a Cottrell
College Science Award from Research Corporation.

%%%%%%%%%%%%%%%%%%%%%%%%%%%%%%%%%%%%%%%%%%%%%%%%%%%%%%%%%%%%%%%%%%%%%%
% Bibliography

%\clearpage
\bibliography{apj-jour,radio}
\bibliographystyle{apj}

%%%%%%%%%%%%%%%%%%%%%%%%%%%%%%%%%%%%%%%%%%%%%%%%%%%%%%%%%%%%%%%%%%%%
\clearpage

% table.fluxes

\begin{center}
% possible sizes are \small(11pt) \footnotesize (10pt) or \scriptsize (8pt)
\tabletypesize{\small} 
%\rotate
\tablewidth{0pt}
\begin{deluxetable}{l r r r r r r }
\tablecaption{Total flux densities of components
\label{table.fluxes} 
}
\tablehead{
\colhead{Component} & \multicolumn{2}{c}{13 cm}         & \multicolumn{2}{c}{18 cm}         & \multicolumn{2}{c}{21 cm} \\
                    & \colhead{Total} & \colhead{RMS}   & \colhead{Total} & \colhead{RMS}   & \colhead{Total} & \colhead{RMS} \\
                    & \colhead{(mJy)} & \colhead{(mJy)} & \colhead{(mJy)} & \colhead{(mJy)} & \colhead{(mJy)} & \colhead{(mJy)} 
}
\startdata
%       13cm           18cm           21cm 
A    & 43.34 & 0.06 & 41.72 & 0.03 & 34.12 & 0.05\\
B    & 25.59 & 0.06 & 23.41 & 0.04 & 22.86 & 0.05\\
G$'$ &  0.83 & 0.07 &  0.53 & 0.03 & 0.46 & 0.05\\
\enddata
\end{deluxetable}
\end{center}

%%%%%%%%%%%%%%%%%%%%%%%%%%%%%%%%%%%%%%%%%%%%%%%%%%%%%%%%%%%%%%%%%%%%
% table.Gpos
%\clearpage
\begin{center}
% possible sizes are \small(11pt) \footnotesize (10pt) or \scriptsize (8pt)
\tabletypesize{\small} 
%\rotate
\tablewidth{0pt}
\begin{deluxetable}{ll ll cl l }
\tablecaption{J2000 positions of the central component relative to image B
\label{table.Gpos} 
}
\tablehead{
\multicolumn{2}{c}{Right Ascension} & \multicolumn{2}{c}{Declination}   & \colhead{Source} & \colhead{Instrument} & \colhead{Reference}\\
\multicolumn{2}{c}{(arcseconds)}    & \multicolumn{2}{c}{(arcseconds)} &                     & 
}
\startdata
0.19   & $\pm$0.03     & 1.00   & $\pm$0.03    & G1   & optical ground  & \citealp{stockton80a} (S80)  \\
0.1820 & $\pm$0.0035   & 1.0178 & $\pm$0.0035  & G1   & optical HST  & \citealp{bernstein97a} (B97) \\
0.155  & $\pm$0.001    & 1.050  & $\pm$0.001   & G    & radio   VLA  & \citealp{roberts85a} (R85) \\
0.16   & $\pm$0.01     & 1.03   & $\pm$0.01    & G    & radio   VLA  & \citealp{harvanek97a}  (H97) \\
0.185  & $\pm$0.001    & 1.028  & $\pm$0.001   & G$'$ & radio   VLBI & \citealp{gorenstein83a} (G83) \\
0.1844 & $\pm$0.0015   & 1.0278 & $\pm$0.0020  & G$'$ & radio   VLBI & this work \\ %Josh 02Dec
\enddata
\end{deluxetable}
\tablecomments{All uncertainties given here are the 1$\sigma$
  uncertainties quoted by the authors.}
\end{center}

%%%%%%%%%%%%%%%%%%%%%%%%%%%%%%%%%%%%%%%%%%%%%%%%%%%%%%%%%%%%%%%%%%%%
%\clearpage

\end{document}